\documentclass[11pt]{article}

\usepackage{times}
\usepackage[T1]{fontenc}
\usepackage[latin1]{inputenc}
\usepackage{geometry}
\usepackage{graphicx,color,rotating,tabularx,dcolumn,amsmath}
\usepackage{pdflscape,longtable}
\usepackage{pdfpages}
\usepackage{soul}  

\usepackage{setspace}

\usepackage[font=doublespacing]{caption}

\makeatletter
\def\@seccntformat#1{\@ifundefined{#1@cntformat}%
   {\csname the#1\endcsname\quad}  
   {\csname #1@cntformat\endcsname}
}
\let\oldappendix\appendix 
\renewcommand\appendix{%
    \oldappendix
    \newcommand{\section@cntformat}{\appendixname~\thesection\quad}
}
\makeatother

\begin{document}

\begin{flushleft}
{\Large
\textbf{Misspecification in mixed-model based association analysis}
}
%
%
\\
Willem Kruijer$^{1}$
\\
\bf{1} Biometris, Wageningen University and Research Centre, Wageningen, Netherlands
\end{flushleft}


\vspace*{0.1in}
\noindent
\textbf{Submitted to \emph{Genetics}}
\vspace*{0.1in}

\noindent
\textbf{Running head}: Misspecification in association analysis 



\vspace*{0.3in}
\noindent
\textbf{*Corresponding author}: \\
Willem Kruijer \\
Biometris \\
Wageningen University and Research Centre \\
PO Box 100, 6700AC Wageningen \\
The Netherlands \\
Phone: +31 317 480806 \\
Email: willem.kruijer@wur.nl \\

\clearpage

\textbf{Abstract:}
Additive genetic variance in natural populations is commonly estimated using mixed models, in which the covariance of the genetic effects is modeled by a genetic similarity matrix derived from a dense set of markers. 
An important but usually implicit assumption is that the presence of any non-additive genetic effect only increases the residual variance, and does not affect estimates of additive genetic variance.  Here we show that this is only true for panels of unrelated individuals. In case there is genetic relatedness, the combination of population structure and epistatic interactions can lead to inflated estimates of additive genetic variance. 

\vspace*{0.4in}
\noindent

Mixed models with random genetic effects
have become an important tool for studying the genetic architecture of complex traits.
The covariance of the genetic effects is assumed to be proportional to a genetic similarity matrix (GSM) based on a dense set of markers, which is equivalent to assuming
additive effects for each standardized marker score. Under several additional assumptions,
such as constant LD,
this gives unbiased estimates of
additive genetic variance and narrow-sense heritability (\cite{yang_etal_2010}, \cite{speed_hemani_johnson_balding_2012}, \cite{speed_balding_2015}).
The sampling variance of such heritability estimators has been studied in \cite{visscher_goddard_2014} and \cite{kruijer_etal_2015}.
These results are however derived under the assumption that the model is correct, i.e. contains the true distribution of the data.
Here we consider situations where this is not the case, and argue that potential sources of bias 
may be identified by computing the parameter value $\tilde \theta$ which minimizes 
the Kullback-Leibler divergence $KL(Q,P_\theta) = \int \log(Q / P_\theta) dQ$ with respect to the true distribution $Q$.
It is a well known fact from statistics that in case of misspecification, i.e. when $Q$ is not contained in the model $\{P_{\theta} : \theta \in \Theta\}$, the maximum likelihood (ML) estimator converges to $\tilde \theta$
(\cite{huber_1967},
\cite{white_1982}). 
Several authors have studied missing or phantom heritability resulting from undetected epistatic interactions between specific loci (\cite{culverhouse_etal_2002}, \cite{song_etal_2010}, \cite{zuk_etal_2012}).
Here we investigate misspecification in a mixed model context, the covariance of the data being misspecified due to infinitesimal interactions or other non-additive effects.
We consider three different scenarios (A-C), each time assuming that the additive and non-additive genetic variance is respectively $0.4$ and $0.2$.
The total phenotypic variance is assumed to be known and equal to $1$, giving a narrow- and broad-sense heritability of $0.4$ and $0.6$.

\textbf{Scenario A:} the phenotype $Y=(Y_1,\ldots,Y_n)' $ of $n$ individuals is modeled using the multivariate normal  distribution 
\begin{equation}\label{m1}
P_{\sigma_A^2,\sigma_E^2}  = N(0,\sigma_A^2 K + \sigma_E^2 I_n),
\end{equation}
where $K$ is a marker-based GSM, $I_n$ the identity matrix, $\sigma_A^2 \in [0,1]$ is the additive genetic variance and $\sigma_E^2=1-\sigma_A^2$ is the residual variance.
We assume however that $Q$, the \emph{actual} distribution of $Y$, is the zero mean normal distribution with covariance $0.4 K + 0.2 (K \cdot K)  + 0.4 I_n$, $\cdot$ being the Hadamard ( entry-wise) product. The 'epistatic' matrix $(K \cdot K)$ is the covariance due to small epistatic interactions between all standardized marker scores (File S1).
Since $(K \cdot K)$ does not equal the identity matrix $I_n$, $Q$ is not contained in model \ref{m1}. Hence, the ML-estimator will not converge to $Q$, but rather to the point $(\tilde \sigma_A^2,\tilde \sigma_E^2)$ minimizing the KL-divergence $KL(Q,P_{\sigma_A^2,\sigma_E^2})$. For genetic similarity matrices derived from published data in maize, rice and Arabidopsis, $\tilde \sigma_A^2$ ranges between $0.47$ and $0.53$ (Table \ref{KL_table}). Hence, the presence of epistatic interactions leads to inflated estimates of additive genetic variance.
For a panel of simulated unrelated individuals, $\tilde \sigma_A^2$ is $0.40$, which is due to the much smaller off-diagonal elements of $K$, making $K \cdot K$ almost indistinguishable from $I_n$.

\textbf{Scenario B:} a plant trait is phenotyped on $r$ genetically identical replicates. Following \cite{kruijer_etal_2015}, the observations $Y=(Y_{11},\ldots,Y_{nr})'$ are modeled by the normal distribution
\begin{equation}\label{m2}
P_{\sigma_A^2,\sigma_E^2} = N(0,\sigma_A^2 Z K Z' + \sigma_E^2 I_{nr}),
\end{equation}
$Z$ being an incidence matrix assigning plants to genotypes. The true distribution $Q$ is multivariate normal with covariance $0.4 Z K Z' + 0.2 Z Z' + 0.4 I_{n r}$, i.e. there are non-additive (not necessarily epistatic) effects 
with independent  $N(0,0.2)$ distributions. Such effects could be due to for example genotype-environment interaction. In contrast to model \ref{m1} (where $Z=I_n$ and $r=1$), $Z Z'$ is different from $I_{nr}$, and $Q$ is not contained in model \ref{m2}.
Again, the value $\tilde \sigma_A^2$ minimizing KL-divergence is substantially larger than $0.4$ (Table \ref{KL_table}), and additive genetic variance will tend to be overestimated. Intuitively, this is because the block structure $Z Z'$ is better captured by $Z K Z'$ than by the diagonal residual.

\textbf{Scenario C} is a combination of A and B. To avoid the misspecification occurring in scenario B, the model
\begin{equation}\label{m3}
P_{\sigma_A^2,\sigma_G^2,\sigma_E^2} =  N(0,\sigma_A^2 Z K Z' + \sigma_G^2 Z Z' + \sigma_E^2 I_N)
\end{equation}
is considered, extending
\eqref{m2} with independent non-additive effects.
This model has been used in the analysis of field trials
(\cite{oakey_etal_2006}, \cite{oakey_etal_2007}), as well as genomic prediction (\cite{gianola_vankaam_2008}, \cite{howard_etal_2014}, \cite{jarquin_etal_2013}).
If in fact the non-additive effects have covariance $K \cdot K$ (as in scenario A), the data have covariance
$0.4 Z K Z' + 0.2 Z (K \cdot K) Z' + 0.4 I_{n r}$. As in scenarios A and B, the $\tilde \sigma_A^2$ minimizing KL-divergence is larger than (Table \ref{KL_table}), while $\tilde \sigma_E^2$ was always $0.40$.

\begin{table}[ht]
\centering
\begin{tabular}{|r|r|r|r|r|r|}
  \hline
Population / source & species & size ($n$) & A & B & C \\
  \hline
Swedish regmap & \emph{A. thaliana} & 298 & 0.53 & 0.58 & 0.53 \\
  Hapmap & \emph{A. thaliana} & 350 & 0.47 & 0.60 & 0.48 \\
Van Heerwaarden \emph{et al.} &   \emph{Z. mays} & 400 & 0.50 & 0.58 & 0.50 \\
Zhao \emph{et al.} & \emph{O. sativa} & 413 & 0.51 & 0.52 & 0.50 \\
Unrelated individuals & simulated & 3000 & 0.40 & & \\
   \hline
\end{tabular}
\caption{
{\bf Values of the additive genetic variance ($\tilde \sigma_A^2$) minimizing the Kullback-Leibler divergence $KL(Q,P)$ with respect to the true distribution ($Q$) of scenarios A-C, with $P$ contained in models \ref{m1}-\ref{m3}.}
Minimization was performed by evaluating KL-divergence on the grid $0,0.01,\ldots,1$ for all variance components, under the constraint they sum to one.
Five populations were considered: the Arabidopsis Hapmap and Swedish regmap (\cite{horton_etal_2012}, \cite{kruijer_etal_2015}), the rice population from \cite{zhao_etal_2011}, the maize population of \cite{vanheerwaarden_etal_2012a} and a simulated population (File S1). Except for the latter, there are $r=2$ replicates of each genotype.
}
\label{KL_table}
\end{table}

In addition to the analysis of KL-divergence we analyzed simulated traits for the first 4 populations, for which we found similar or even larger bias (File S2).
This has important implications,
in particular for immortal populations, for which genetically identical replicates are available (e.g. \emph{A. thaliana}, agronomic crops, bacteria and fungi). Typically there is strong population structure and often only several hundreds of different genotypes are phenotyped. One can analyze such data at individual level (model \ref{m2}) or at the level of genotypic means (model \ref{m1}, with $\sigma_E^2$ divided by the number of replicates). \cite{kruijer_etal_2015} showed that in the latter type of analysis, standard errors of heritability 
estimates can be huge, and recommended model \ref{m2} for both heritability estimation and genomic prediction. Here we have shown that in presence of non-additive effects, this model is likely to overestimate additive genetic variance. If however the non-additive effects are due to epistatic interactions, analysis at genotypic means level (model \ref{m1}) will (apart from the large sampling variance) also give inflated estimates of additive genetic variance. This is a rather realistic scenario, since epistasis may be an important part of the genetic architecture (\cite{mackay_2014}), and several other types of non-additive effects can be ruled out or minimized for immortal populations: e.g. genotype by environment interactions are unlikely in homogeneous controlled environments with adequate randomization, and dominance effects are impossible when using inbred lines. 

Interestingly, 
the inflation of additive genetic variance 
is not due to any non-linearity or absence of main effects, but rather the population structure present in the epistatic GSM, which to some extent resembles the structure of the GSM for the additive effects. At the same time, it is this structure that makes the epistatic GSM distinguishable from the diagonal error. 
This suggests that epistatic interactions are easier to model in structured populations, i.e. sampling variance of epistatic variance components may not be as large as in unstructured human populations (\cite{Yang_HongLee_Goddard_Visscher_2011}). Expressions for the asymptotic variance in a model with both additive and epistatic effects (File S3) indicate that this is indeed the case.
More generally, the inflation of heritability estimates due to misspecification 
illustrates the difficulty of modeling and estimating genetic effects. As recently pointed out by \cite{speed_balding_2015} this is already challenging for the additive genetic effects, in the sense that depending on the genetic architecture different GSMs may be appropriate. Indeed, the potential bias resulting from an inappropriate GSM could be assessed by evaluating KL-divergence with respect to the true model, as is the case for alternatives for the epistatic GSM 
considered here.


\section*{Acknowledgments}

Martin Boer and Fred van Eeuwijk are acknowledged for useful comments on the manuscript.
The research leading to these results has been conducted as part of the DROPS project which received funding from the European Community's Seventh Framework Programme (FP7/ 2007-2013) under the grant agreement number 244374. The research was also funded by the Learning from Nature project of the
Dutch Technology Foundation (STW), which is part of the
Netherlands Organisation for Scientific Research (NWO).

%
%
%
%

\renewcommand\refname{Literature Cited}
\bibliography{d:/willem/research/statgen3}

\end{document}